\pdfoutput=1
\documentclass{svjour3}                     

\smartqed  

\usepackage{graphicx,lineno,bm,float,soul,amsfonts,amsmath,siunitx}
\usepackage{xcolor,hyperref}

\modulolinenumbers[5]

\newcommand{\T}{^{\mbox{\tiny T}}}

\newcommand{\B}[1]{{\bm #1}}

\newcommand{\dd}{\; \text{d}}
\newcommand{\ds}{\displaystyle}

\begin{document}

\title{Orbit transfer using Theory of Functional Connections via change of variables}

\author{Allan K. de Almeida Jr $\dagger$* \and Antonio F. B. A. Prado* \and Daniele Mortari$\ddagger$}

\institute{$\dagger$ CICGE, Faculdade de Ciências da Universidade do Porto, 4169-007 Porto, Portugal \\
*
	INPE - National Institute for Space Research, S\~ao Jos\'e dos Campos, SP, Brazil \\
	$\ddagger$ Aerospace Engineering, Texas A\&M University, College Station TX, 77843. \\
	$\dagger$ \email{allan.junior@ua.pt}\\
	* \email{antonio.prado@inpe.br}\\
	$\ddagger$ \email{mortari@tamu.edu} \\
}

\date{Paper submitted to The European Physical Journal Special Topics}

\maketitle

\begin{abstract}
    This work shows that a class of astrodynamics problems subject to mission constraints can be efficiently solved using the Theory of Functional Connections (TFC) mathematical framework by a specific change of coordinates. In these problems, the constraints are initially written in non-linear and coupled mathematical forms using classical rectangular coordinates. The symmetries of the constrained problem are used to select a new system of coordinates that transforms the non-linear constraints into linear. 
    This change of coordinates is also used to isolate the components of the constraints.
    This way the TFC technique can be used to solve the ordinary differential equations governing orbit transfer problems subject to mission constraints. Specifically, this paper shows how to apply the change of coordinates method to the perturbed Hohmann-type and the one-tangent burn transfer problems.
\end{abstract}


\section{Introduction}

The Theory of Functional Connections (TFC) is a new important and recent mathematical framework \cite{TFC_Book,U-ToC} that provides efficient methods to solve a class of constrained optimization problems.
This mathematical tool allows to solve constrained optimization problems, such as Boundary Value Problems (BVP), using unconstrained differential equations.
This is done by deriving \textit{constrained expressions}, which are functionals with embedded constraints. These functionals reduce the whole function space (where to search the solution) to just the function subspace that fully satisfy the problem constraints. In this way, the initial constrained problem is transformed into an unconstrained problem that can be solved using more simple, robust, accurate, fast and reliable methods. As an example, a BVP on a linear ODE is directly solved by linear least-squares.

The most common application of TFC is to solve linear \cite{LDE} and nonlinear \cite{NDE} ODEs, by deriving constrained expressions subject to a variety of constraints, such as points, derivatives, integral, and component constrains, as well as any linear combinations of them. TFC has also been applied to solve hybrid systems \cite{hybrid}, ODEs subject to integral constraints \cite{mca26030065}, integro-differential equations \cite{mca26030065}, eight order BVP \cite{math8030397}, and to solve boundary value geodesic trajectories on curved surfaces \cite{geodesic}.

Although the TFC framework is relatively new, its use to solve the BVP has already been proven efficient when applied to astrodynamics problems \cite{Earth2Moon,periodic,allansr}. When using rectangular coordinates on orbit transfers the constraints equations are often nonlinear (e.g., constraints on position and velocity magnitudes) and coupled. Up to date, TFC has not been used to solve ODEs subject to nonlinear constraints. This paper shows that a change of coordinates can be made to transform the nonlinear and coupled constraints into linear and uncoupled ones. This way, the current TFC framework can be applied to solve transfer problems under the proposed constraints.

The change of coordinates to apply is a function of the types of the constraints and the selected frame of reference. In this paper, it is shown that certain types of constraints and frames of reference show some symmetries. The use of polar coordinates can take the advantage of these symmetries and, consequently, the constraints are described in a mathematical form suitable for TFC. These changes of coordinates allow to benefit from traditional TFC techniques to evaluate transfers. Applications are done for transfers between orbits with different altitudes and perturbed by the Moon under the constraints of Hohmann and one-tangent burn transfer methods \cite{vallado7}. These transfers are evaluated using the circular restricted three body problem (CR3BP) with the perturbation due to the Moon and the two body problem without the Moon. The numerical results allow to quantify the effect of the Moon gravity by comparing the CR3BP and the 2-body results. The method is applied to solve the one-tangent burn transfer, but other types of transfers - constrained in both position and velocity - can be solved using polar coordinates, still taking advantage of the problem's symmetries.

Efficiency is a goal of any maneuver. Depending on the type of transfer, a possible solution related to a maximum efficiency of a burn may be reached at the apogee or perigee of elliptical orbits for a flight-path angle equals to zero \cite{vallado7} (sec. 6), although other symmetrical solutions can be obtained for transfers between co-planar elliptical orbits \cite{MARCHAL196991}.
The condition of a burn at the apogee or perigee is satisfied by the constraint of the component of the velocity (in polar coordinates centered in the massive body) in the radial direction equals to zero, i.e. $\dot{r}=0$.
It is shown in this paper that TFC can be used to analytically embed this linear constraint into the constrained expressions. 
Any solution (numerically obtained) satisfies the above mentioned condition, which may be related to a maximum efficiency, depending on the type of the desired transfer.

Section \ref{id} defines the problem to be solved along with the needed mathematical tools. The transfer constraints and the new method to solve the problem are shown in section \ref{polartfc}. Numerical results for the proposed one-tangent burn method are described in section in section \ref{re}. Finally, the conclusions are written in the last section.

\section{Introduction and definitions}\label{id}

Before presenting the astrodynamic transfer problem, a short summary of TFC is provided.

\subsection{The Theory of Functional Connections}\label{tfc}

In this section, the univariate TFC technique used to solve the BVP is briefly summarized, with the required conditions to be applied. Essentially, the TFC technique reduces the space of all possible functions to its subspace in which the functions must satisfy the constraints of the problem. The TFC involves a functional interpolation in which the constraints of the problem are analytically embedded in an expression called the \textit{constrained expression}. The general equation to derive the \textit{constrained expression} comes from
\begin{eqnarray}\label{eq:ce0}
	x (t, g (t)) = g (t) + \ds\sum_{k = 1}^n \eta_k (t, g (t)) \, s_k (t)
\end{eqnarray}
where $n$ is the number of constraints, $g(t)$ is the free function, $s_k(t)$ constitute a set of $n$ support functions, which must be linearly independent, and $\eta_k (t,g(t))$ are unknown functional coefficients. The unknown functional coefficients are determined using Eq.~(\ref{eq:ce0}) subject to the constraints of the problem under consideration. After that, they are substituted in Eq.~(\ref{eq:ce0}) to form the \textit{constrained expression}, in which all the constraints of the astrodynamics (physical) problem are mathematically included.

The dependent variable is substituted into the differential equation given by the later defined equation of motion according to the \textit{constrained expression}.
After this, a new differential equation arises written in terms of $g(t)$, instead of $x(t)$. Note that this new differential equation is not subject to any constraint. The free function $g(t)$ is then expressed as a linear combination of a set of basis function given by orthogonal polynomials:
\begin{eqnarray}\label{eq:ce1}
	g (t) = \ds\sum_{j = 0}^m \xi_j \, h_j (t) = \B{\xi}\T \B{h} (t) 
\end{eqnarray}
where $m$ is the number of basis functions, $\xi_j$ (for $j = 0, 1, \cdots, m$) are unknown coefficients, and $h_j$ (for $j = 0, 1, \cdots, m$) are orthogonal polynomials. In general, each kind of orthogonal polynomials has intrinsic influence on the numerical evaluations. Although the Chebyshev polynomials are adopted in this research, the Legendre polynomials is also tested, and no significant differences is observed, as will be seen later in the results. Both the Chebyshev and Legendre polynomials are defined only in the range of $t\in [-1,1]$, hence an appropriate change of units of time must be done accordingly. In order to solve the problem, it is enough to find the unknown coefficients $\xi_j$ (for $j = 0, 1, \cdots, m$). Hence, the new differential equation is discretized for a set of $N$ values of time $t$. The best points of $t$ to be chosen is to use Chebychev-Gauss-Lobato nodes \cite{Collo1}, which can be evaluated as
\begin{eqnarray}\label{eq:ce2}
	t_i =-\cos\Big(\dfrac{i \pi}{N}\Big) \qquad \text{for}~ i = 0,1,\cdots,N
\end{eqnarray}
The system of $N$ equations with $m$ unknowns is then solved for the $\xi_j$ (for $j=0,1,\cdots,N$) coefficients using the nonlinear least-squares numerical method. The iterative process is written in Python language, and it uses an automatic differentiation and just-in-time (JIT) compiler \cite{JaxGithub,JaxOriginalPaper,tfc2021github}. More details of the TFC technique to solve ODE can be found in \cite{U-ToC,NDE,TFC-multi-tpa,DEALMEIDA2023102068}.

\subsection{Types of constraints}

The TFC framework can be used for the following types of constraints \cite{U-ToC,mca26030065}:
\begin{itemize}
 \item A $n$-th derivative constraint at a given time $t_0$, which can be expressed by $\dfrac{\dd^k x}{\dd t^k} \Big|_{t_0} = x^{(k)}_0$. 
 \item Two constraints at a given time, which are expressed by $\dfrac{\dd^k x}{\dd t^k}\Big|_{t_0}=x^{(k)}_0$ and $\dfrac{\dd^jx}{\dd t^j}\Big|_{t_0}=x^{(j)}_0$, where $0\le k<j$. Note that the initial value problem for ODE is included here for $k=0$ (initial position) and $j=1$ (initial velocity).
 \item Two constraints at two given times, which are expressed by $\dfrac{\dd^kx}{\dd t^k}\Big|_{t_0}=x^{(k)}_0$ and $\dfrac{\dd^jx}{\dd t^j}\Big|_{t_f}=x^{(j)}_f$. Note that the two point boundary value problem is included here for $k=j=0$.
 \item Relative constraints are also included in the TFC framework, which can be expressed as $\dfrac{\dd^ix}{\dd t^i}\big|_{t_j}=\dfrac{\dd^kx}{\dd t^k}\big|_{t_l}$, where if $i=k$, then $j\ne l$ and if $j=l$, then $i\ne k$.
 \item Any linear combination of the relative constraints is also included in the TFC framework.
 \item ODE subject to an integral constraint, which can be expressed as
 $\ds\int_a^b x (t) \dd t = I$, where $a$, $b$, and $I$ are parameters of the problem. In addition to the integral constraint, the ODE can also be subject to the other linear constraints shown above. For instance, the constrained expression is obtained in \cite{mca26030065} for the following three constraints: $x (t_0) = x_0$, $x (t_f) = x_f$, and $\ds\int_{t_0}^{t_f} x (\tau) \dd \tau = I$.
\end{itemize}




Note that the TFC framework is fully developed for linear constraints. Therefore, its technique to solve ODE is straightforward in the case where the constraints of the problem are linear on its variables.
On the other side, the \textit{constrained expression} may be hard to be obtained in the case where the constraints are nonlinear in the variables.
For instance, the solutions for the functional coefficients $\eta_k$ obtained from the equations applied at the constraints may be not unique in the case where the constraints are nonlinear in the variables. 
This paper proposes to make a transformation of coordinates to avoid this problem. The new coordinates must be such that Eq.~(\ref{eq:ce0}) applied at the constraints generate a set of equations in which the coefficients $\eta_k$ are linear. Note that the solutions for $\eta_k$ are uniquely guaranteed if Eq.~(\ref{eq:ce0}) applied to the constraints generate a set of (linearly independent) equations in which the coefficients $\eta_k$ are linear, because this set constitutes a linear system in which the number of equations is equal the number of variables.
Furthermore, the TFC technique may be complicate to be used for multidimensional problems subject to constraints in which the variables are coupled, i.e. in the case where one dependent variable depends on the value of another dependent variable. For instance, a kind of constraint where $x(t)$ depends on the values of $y(t)$ at a specified time.

As will be shown later, there are several types of constraints which are nonlinear when written in traditional rectangular variables. In order to avoid this problem, a change of coordinates is proposed in this paper. It takes advantage of the symmetry of the system to transform nonlinear constraints into linear ones. In this way, the TFC traditional, robust, accurate, and efficient framework can be used to obtain solutions for the ODE associated to the problem. It can be very useful, in particular for orbit transfers.




\section{The constraints obtained from the Hohmann and one-tangent burn transfer method}\label{polartfc}

The problem of transfer a spacecraft from a circular orbit of radius $r_0$ to another co-planar circular orbit of radius $r_f$ using a bi-impulsive maneuver is analyzed next. One way to make the transfer is the Hohmann transfer, which determines a specific elliptical orbit for the spacecraft during the transfer. The lowest $\Delta V$ is reached if the impulses are applied at the perigee and apogee of the elliptical transfer orbit in the direction tangent to the velocity at those points. The time required for this transfer is $T_h = (\pi/2) \, \sqrt{(r_0 + r_f)^3/\mu}$, where $\mu$ is the gravitational parameter of the massive body.
An alternative to the Hohmann transfer is the one-tangent burn \cite{vallado7}, which is also given by two impulses, but with only one of them is tangent to the trajectory. The one-tangent burn method has the advantage of a variable time of flight (smaller or larger), at the cost of increasing the fuel consumption, since only one of the impulses is required to be tangent to the trajectory. The trade-off between cost and time of flight offered by the one-tangent burn method can be beneficial, according to the type of the mission.
Note that the one-tangent burn method is a transfer of a satellite from an initial circular orbit of radius $r_0$ to a final orbit of radius $r_f$ in a total time of flight $T$. In the particular case, where $T=T_h$, the one-tangent burn coincides with the Hohmann transfer.
The trajectories for this type of transfer are shown in Fig. \ref{fig:f1}, for different values of the time of flight, in the case where there is no other perturbative force than the gravitational attraction of the primary.
\begin{figure}[htbp]\centering
	\includegraphics[scale=0.4]{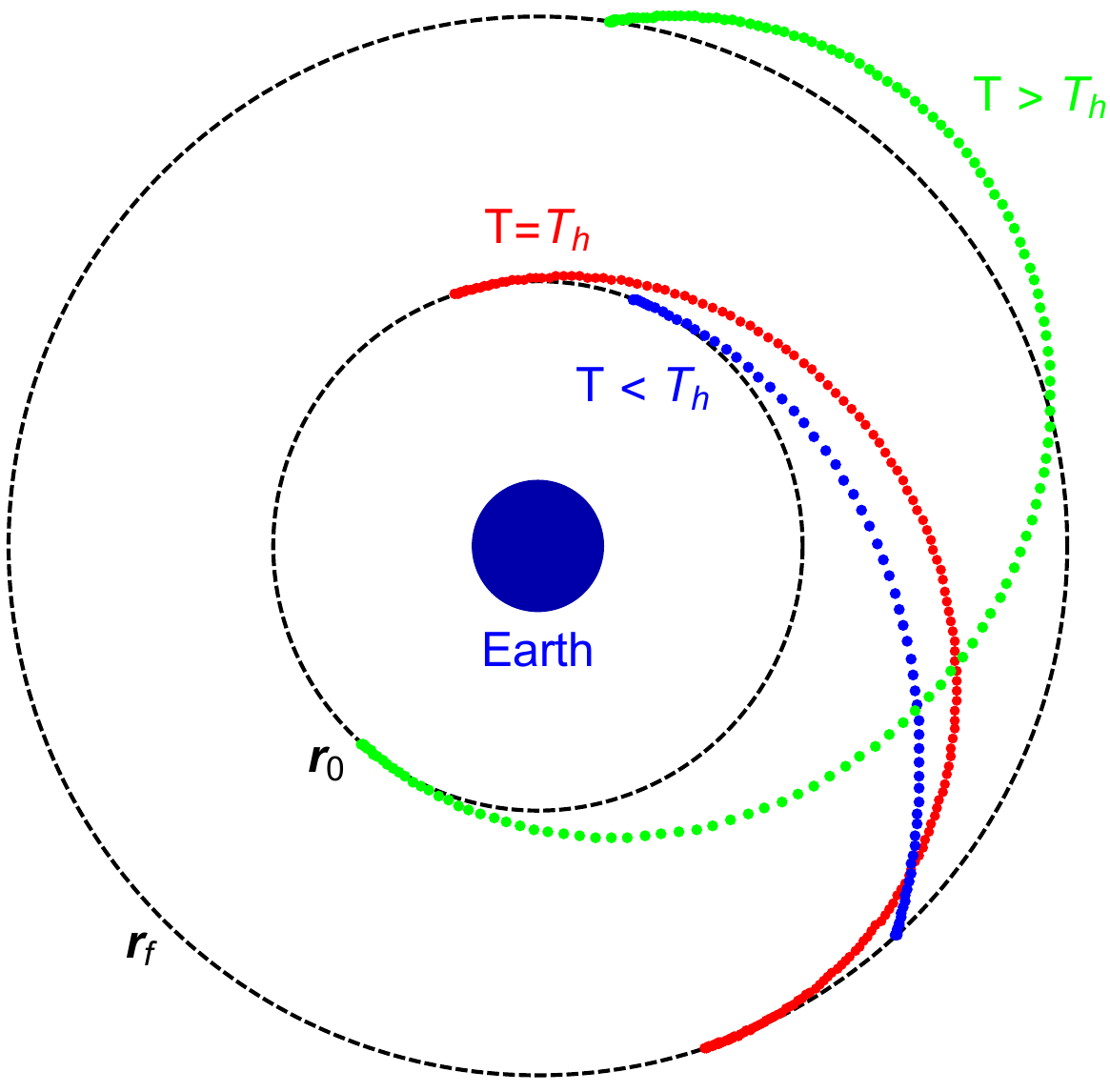}
	\caption{One-tangent burn transfer trajectories from the initial circular orbit with radius $r_0$ to the final circular orbit with radius $r_f$ for different values of time of flight. The trajectory in red is coincident with a Hohmann transfer. The time of transfer of the blue trajectory is lower than the time of the Hohmann transfer, while the time of transfer of the red is larger.}
	\label{fig:f1}
\end{figure}
The one-tangent burn transfer method is subject to the following constraints:
\renewcommand{\theenumi}{\roman{enumi}}%
\begin{enumerate}
	\item The distance of the position to the center of the frame of reference is $r_0$ when $t = 0$.
	\item The distance of the position to the center of the frame of reference is $r_f$ when $t = T$.
	\item The velocity is pointed to the direction tangential to the trajectory when $t = 0$.
\end{enumerate}
Note that although the constraint iii is not a constraint of the formulation of the Hohmann transfer problem, it is a result of the optimization procedure. Thus, if the constraint iii is considered in the transfer, the optimized solution given by the Hohmann transfer can be retrieved by the TFC optimization procedure in the case where $T=T_h$.
In a frame of reference centered in the center of the Earth, the constraints i, ii, and iii are written in traditional rectangular coordinates as
\begin{enumerate}
	\item $\sqrt{x^2 + y^2}|_{t = 0} = r_0$
	\item $\sqrt{x^2 + y^2}|_{t = T} = r_f$
	\item $(\dot{x} \, x + \dot{y} \, y)|_{t = 0} = 0$
\end{enumerate}
These constraints, written in rectangular coordinates, are in nonlinear and coupled (the variable $x$ depends on the variable $y$ for some times) mathematical forms. TFC framework has been fully developed for linear constraints.
As explained in subsection \ref{tfc}, the TFC technique has not been used to solve differential equations under these forms of constraints.
On the other side, it is shown next that this issue is avoided by writing these constraints after a convenient change of coordinates. Using polar coordinates $(r,\theta)$, the constraints i, ii, and iii can be expressed as
\begin{enumerate}
	\item $r|_{t = 0} = r_0$
	\item $r|_{t = T} = r_f$
	\item $\dot{r}|_{t = 0} = 0$
\end{enumerate}
Now, the expressions for the constraints i, ii, and iii are linear and uncoupled (the variable $r$ does not depend on the variable $\theta$). This means that TFC can be applied, as it has been developed, to solve the equations of motion associated to the constrained transfer problem. 

\subsection{Analyses of the transfer method}


The constraints i, ii, and iii shown above are the same ones for the Hohmann and the one-tangent burn methods of transfer, but it is important to note that no restriction was imposed into the equations of motion.
The solution to the transfer must satisfy the constraints obtained from the 2-body problem shown above in section \ref{polartfc} (obtained from the Hohmann and the one-tangent burn methods) even in the case that the satellite is subject to several perturbative forces other than the central gravitational attraction of the primary.
Thus, the method shown in this paper is general, since it is not restricted to the 2 body-problem, i.e. perturbations can be included in the equations of motion.
In fact, transfers subject to these constraints will be done later in this paper for a satellite under the perturbation due to the gravitational attraction of the Moon, in addition to the Earth.

The choice of the system of coordinates - which transform nonlinear constraints into linear ones - depends on both the constraints of the mission and the frame of reference.
In general, a frame of reference centered in one of the primaries shows symmetries with orbits also centered in this primary. The use of polar coordinates can take the advantage of these symmetries.
Depending on the type of transfer and the initial and final orbits, the lowest delta-$v$ for impulsive maneuvers is reached when the impulse is applied at the apogee or perigee in the tangential direction of an orbit.
The mathematical form of this constraint is simply $\dot{r}=0$ (a linear form) in polar coordinates centered in the primary. Thus, the solutions obtained via TFC will always satisfy this useful constraint. The change of coordinates procedure shown in this paper can also be used with constraints other than the ones showed in the previous subsection to take the advantage of such efficiency, according to the objectives of the mission.

\subsection{Obtaining the \textit{constrained expression} for the one-tangent burn constraints in polar coordinates}

Using the polar coordinates shown above, the two dependent variables are $r (t)$ and $\theta (t)$. Using Eq.~(\ref{eq:ce0}) for $r(t)$, instead of the dependent variable $x (t)$, the general equation used to derive the \textit{constrained expression} is 
\begin{eqnarray}
	\label{eq:cer0}
	r (t, g_r (t)) = g_r (t) + \eta_1 \, s_1 (t) + \eta_2 \, s_2 (t) + \eta_3 \, s_3 (t)
\end{eqnarray}
where $g_r (t)$ is the free function, $\eta_1$, $\eta_2$, and $\eta_3$ are constant coefficients, and $s_1 (t)$, $s_2 (t)$, and $s_3(t)$ are the support functions.
The constraints i, ii, and iii can be written as
\begin{eqnarray}
	\label{eq:cer1}
	r (0, g_r (0)) = r_0 \\ \nonumber
	r (T, g_r (T)) = r_f \\ \nonumber
	\dot{r} (0, g_r (0)) = 0
\end{eqnarray}
Using Eq.~(\ref{eq:cer0}), these constraints become
\begin{eqnarray}
	\label{eq:cer2}
	r_0 = g_r (0) + \eta_1 \, s_1 (0) + \eta_2 \, s_2 (0) + \eta_3 \, s_3 (0)\\ \nonumber
	r_f = g_r (T) + \eta_1 \, s_1 (T) + \eta_2 \, s_2 (T) + \eta_3 \, s_3 (T)\\ \nonumber
	0 = \dot{g}_r (0) + \eta_1 \, \dot{s}_1 (0) + \eta_2 \, \dot{s}_2 (0) + \eta_3 \, \dot{s}_3 (0)
\end{eqnarray}
The support functions are given by the first three terms of the Chebyshev polynomials: $s_1 (t)=1$, $s_2 (t)=t$, and $s_3 (t)=-1+2t^2$. Thus, Eqs~(\ref{eq:cer2}) become
\begin{eqnarray}
	\label{eq:cer3}
	r_0 &=& g_r (0) + \eta_1  -  \eta_3 \\ \nonumber
	r_f &=& g_r (T) + \eta_1  + \eta_2 \, T + \eta_3 \, (-1+2T^2)\\ \nonumber
	0 &=& \dot{g}_r (0) +  \eta_2  
\end{eqnarray}
Solving the system of three equations for the three linear coefficients ($\eta_1$, $\eta_2$, and $\eta_3$) given by Eqs.~(\ref{eq:cer3}) and substituting them into Eq.~(\ref{eq:cer0}), the \textit{constrained expression} for the $r$ variable is written as 
\begin{align}\label{eq:consttrainedexpression}
	r (t, g_r (t)) &= g_r (t) + r_0 - g_r(0) - t \, \dot{g}_r (0)  \nonumber \\ ~&~ \; + \dfrac{t^2}{T^2} \left[\dot{g}_r (0) \, T + g_r(0) - g_r (T) - r_0 + r_f\right] 
\end{align}
Note that $r (t, g_r (t))$ given by Eq.~(\ref{eq:consttrainedexpression}) always satisfies the constraints i, ii, and iii, independently of the form of the free function $g_r (t)$.

The acceleration due to the gravitational attraction of the Earth is central (does not depend on the angle) and the transfer is done between circular orbits, thus the cost of the transfer using the two-body problem is independent of the position along the orbit in which the first impulse is applied. On the other side, this is not true when the gravity of the Moon is taken into account by using the CR3BP, as will be shown in the next subsection. Thus, in order to facilitate the evaluations, the transfer is additionally subject to the following constraint: $\theta(0) = \theta_0$, where $\theta_0$ is a given value that defines the position along the orbit where the first impulse is applied.
Thus, using $\theta(t)$ instead of $x(t)$, Eq.~(\ref{eq:ce0}) becomes
\begin{eqnarray}\label{eq:ceth0}
	\theta(t, g_\theta (t)) = g_{\theta} (t) + \eta_{\theta} \, s_1(t)
\end{eqnarray}
where $\eta_{\theta}$ is obtained from the constraint $\theta (0) = \theta_0$. Hence, the \textit{constrained expression} becomes
\begin{eqnarray}\label{eq:consttrainedexpressionth}
	\theta(t, g_{\theta}(t)) = g_{\theta} (t) - g_{\theta} (0) + \theta_0
\end{eqnarray}
where $g_{\theta} (t)$ is the free function.






\subsection{The Circular Restricted Three Body Problem in polar coordinates}

A frame of reference is rotating with angular velocity $\B{\omega}$ with respect to an inertial frame with a shared common center. The acceleration of a body in this inertial frame of reference with respect to the acceleration, velocity, and position in the rotating frame of reference is
\begin{eqnarray}\label{eq:accrot}
	\frac{\dd^{*2} \B{r}^{*}}{\dd t^2} = \frac{\dd^{2} \B{r}}{\dd t^2} + 2 \, \B{\omega} \times \frac{\dd \B{r}}{\dd t} + \B{\omega} \times \left(\B{\omega} \times \B{r}\right) + \frac{\dd^{*} \B{\omega}}{\dd t} \times \B{r}
\end{eqnarray}
where $\B{r}^{*}$ and $\B{r}$ are the positions with respect to the inertial and rotating frames, respectively. The derivatives with no star must be taken in the rotating frame (not considering the motion of its bases with respect to the inertial frame), while the derivatives with star are taken with respect to the inertial frame. Equation \ref{eq:accrot} shows the terms required to obtain the equations of motion in the rotating frame of reference from the Newton's second law. Details can be seen in, e.g. \cite[ch. 7]{symon}.

The above defined inertial frame of reference is centered in the Earth, while the rotating frame of reference rotates with the same constant angular velocity of the circular motion of the Moon around the Earth. The rotating frame of reference and the positions of the bodies involved are shown in Fig.~\ref{fig:f1}. Since the angular velocity is constant, the last term of Eq.~(\ref{eq:accrot}) is zero. Using the acceleration in the inertial frame given by Eq.~(\ref{eq:accrot}), the equation of motion of a satellite in the rotating frame of reference described above becomes
\begin{eqnarray}\label{eq:crtbp}
	\frac{\dd^2 \B{r}}{\dd t^2} + 2 \, \B{\omega} \times \frac{\dd \B{r}}{\dd t} + \B{\omega} \times \left(\B{\omega} \times \B{r}\right) =- \dfrac{\mu_e}{r^3} \, \B{r} - \dfrac{\mu_m}{r_m^3} \, \B{r}_m
\end{eqnarray}
where $\B{r}$ is the position of the satellite, $r$ is the magnitude of $\B{r}$, $\B{r}_m$ is the position of the satellite with respect to the Moon, $r_m$ is the magnitude of $\B{r}_m$, $\mu_e$ and $\mu_m$ are the gravitational parameter of the Earth and Moon, respectively, and $\B{\omega}=(0,0,\omega)$ is the angular velocity of the rotating frame of reference, where the angular speed is $\omega=\sqrt{\mu_m/R^3}$, where $R$ is the distance between the Earth and the Moon. Note that the derivatives with no star are taken with respect to the rotating frame of reference - the derivatives (with no star) of its basis with respect to time are equal to zero. Although Eq.~(\ref{eq:crtbp}) describes the CR3BP, the rotating frame coincides with the inertial one in the case where $\mu_m=0$, hence, $\omega=0$, a special case representing the two body problem.
\begin{figure}[htbp]
\centering
	\includegraphics[scale=0.3]{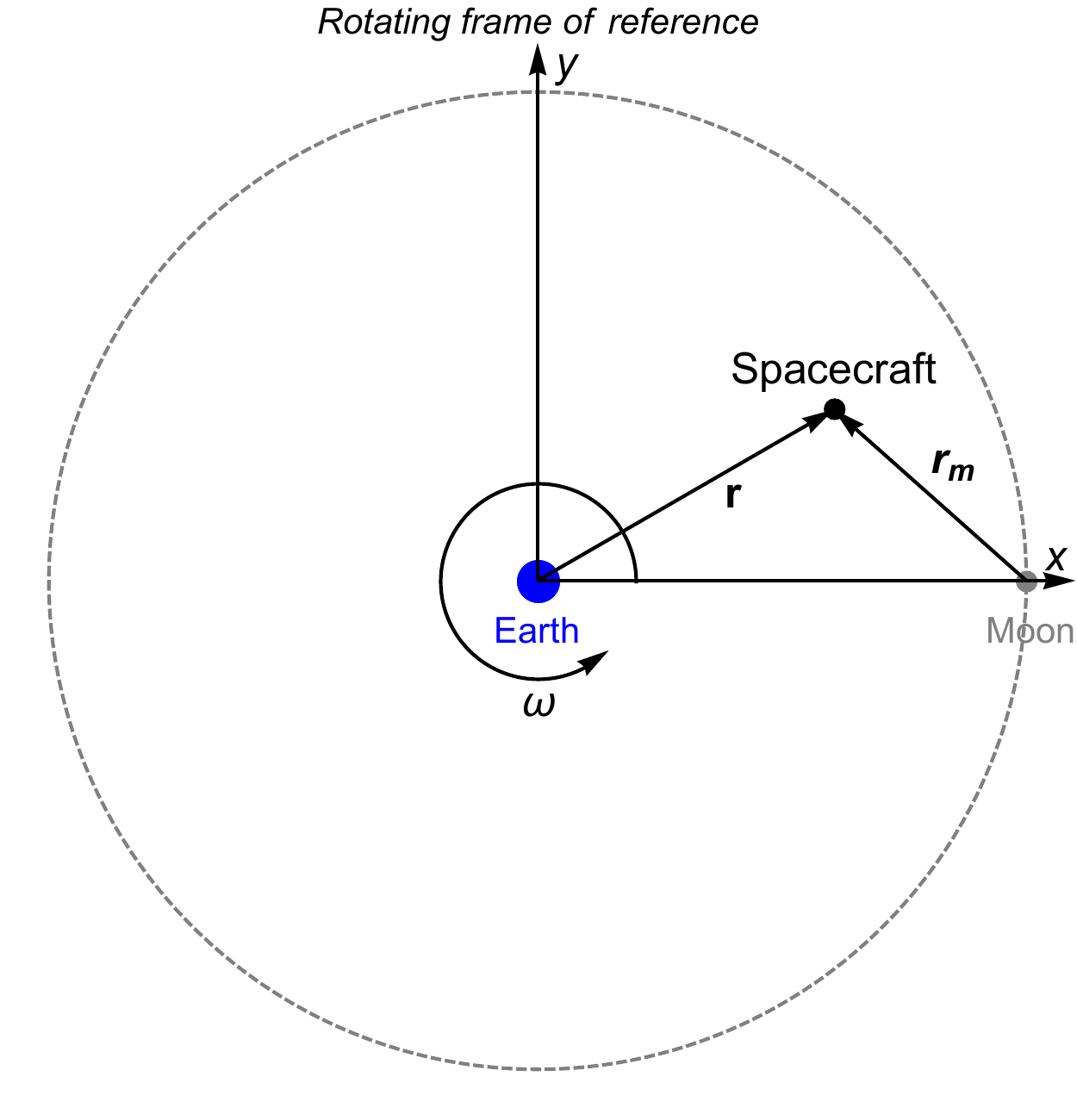}
	\caption{The rotating frame of reference centered in the Earth.}
	\label{fig:f1}
\end{figure}

\begin{figure}[htbp]
\centering
    \includegraphics[scale=0.3]{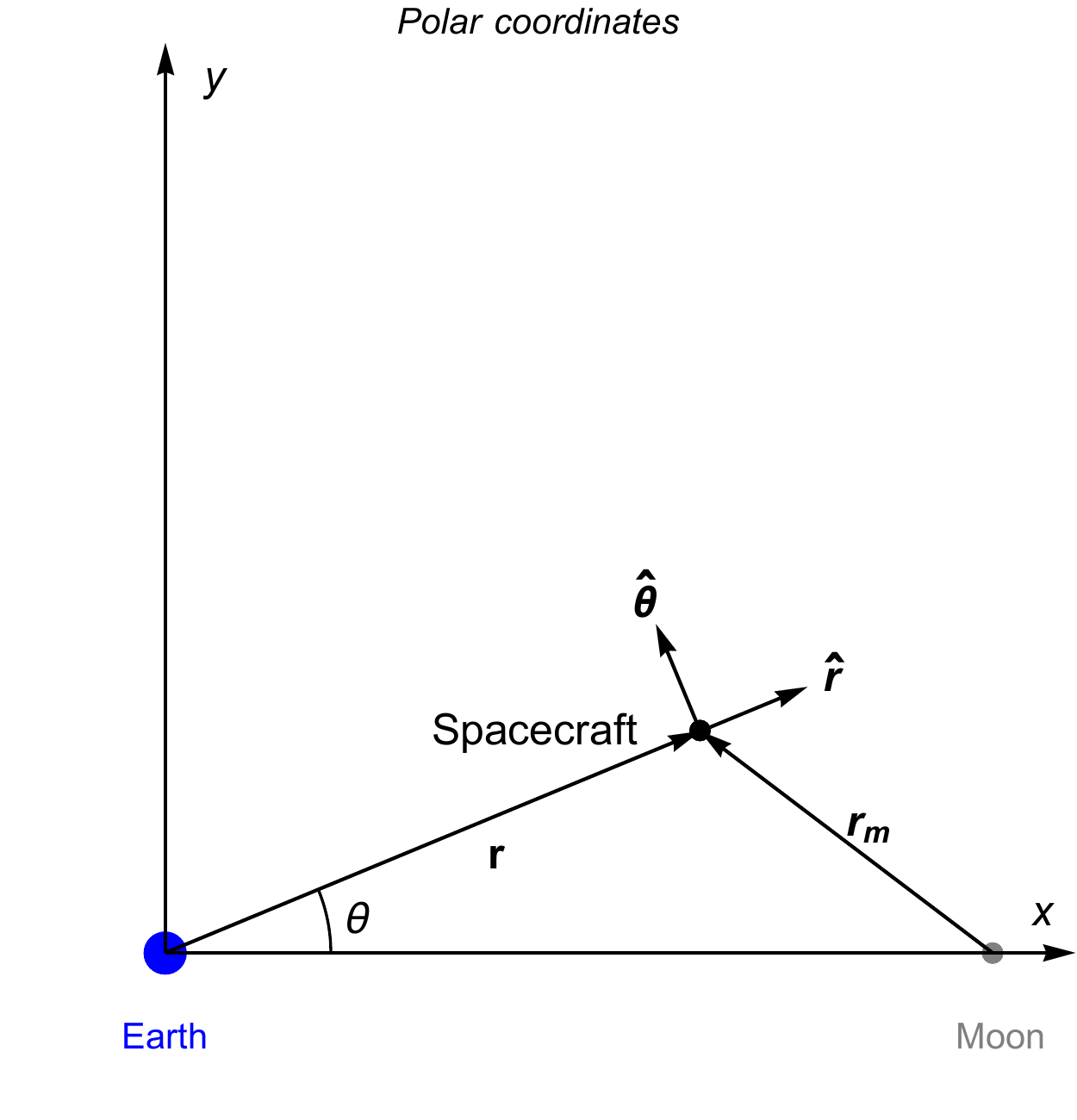}
	\caption{Polar coordinates in the rotating frame of reference.}
	\label{fig:f2}
\end{figure}

The position of the satellite in the rotating frame of reference can be written using standard rectangular coordinate system as
\begin{eqnarray}\label{eq:reference1}
	\B{r} = x \, \B{i} + y \, \B{j}
\end{eqnarray}
where $\B{i}$ and $\B{j}$ are the two unit-vector along associated with the $x$ and $y$ coordinates, respectively, in the rotating frame. Alternatively, traditional polar coordinates can also be used in the rotating frame of reference by defining two unit vectors $(\hat{r}, \hat{\theta})$ from the standard rectangular coordinates $(x, y)$ according to
\begin{equation}\label{eq:referencer}
	\begin{cases} \hat{\B{r}} = \cos\theta \, \B{i} + \sin\theta \, \B{j} \\ \hat{\B{\theta}} =-\sin\theta \, \B{i} + \cos\theta \, \B{j}\end{cases}
\end{equation}
A draw of the rectangular and polar coordinates in the rotating frame is shown in Fig.~\ref{fig:f2}.
Thus, the position, velocity, and acceleration are given by
\begin{equation}\label{eq:pos}
	\begin{cases} \B{r} = r \, \hat{\B{r}}, \\ 
		\dot{\B{r}} = \dot{r} \, \hat{\B{r}} + r \, \dot{\theta} \, \hat{\B{\theta}}, \\
		\ddot{\B{r}} = \left(\ddot{r} - r \, \dot{\theta}^2\right) \hat{\B{r}} + \left(\dfrac{1}{r} \dfrac{\dd}{\dd t} \left(r^2 \, \dot{\theta}\right)\right) \hat{\B{\theta}}\end{cases}
\end{equation}
It is important to note that the dots represent the derivatives with no star with respect to time (in the rotating frame).
Using the polar coordinates defined above, the second and third terms on the left side of Eq.~(\ref{eq:crtbp}) become
\begin{eqnarray}\label{eq:po1}
	\begin{cases}
		2 \, \B{\omega} \times \dot{\B{r}}=-2 \omega r \dot{\theta} \hat{\B{r}}+2\omega \dot{r}\hat{\B{\theta}} \\ \nonumber
		\B{\omega} \times \left(\B{\omega} \times \B{r}\right)=-\omega^2 r	 \hat{\B{r}}
	\end{cases}
\end{eqnarray}
The gravitational acceleration due to the Earth in the ecliptic of the Earth-Moon is given by
\begin{eqnarray}\label{eq:po2}
	- \dfrac{\mu_e}{r^3} \, \B{r} &=&  - \dfrac{\sqrt{r^2} \mu_e}{r^3} \hat{\B{r}}
\end{eqnarray}
and the gravitational acceleration due to the Moon is
\begin{eqnarray}\label{eq:po3}
	- \dfrac{\mu_m}{r_m^3} \, \B{r}_m &=& \dfrac{\mu_m (R \cos (\theta)-r)}{\sqrt{r^2-2 r R \cos (\theta)+R^2}}\hat{\B{r}} -\dfrac{\mu_m R \sin (\theta)}{\sqrt{r^2-2 r R \cos (\theta)+R^2}} \hat{\B{\theta}}
\end{eqnarray}
Hence, the equations of motion given by Eq.~(\ref{eq:crtbp}) can be written using polar coordinates as
\begin{gather}\label{eq:movpol}
	\ddot{r} - r \, \dot{\theta}^2 - r \omega^2 - 2 r \, \omega \, \dot{\theta} + \dfrac{\sqrt{r^2} \mu_e}{r^3} 
	+ \dfrac{(r \mu_m-R \mu_m \cos\theta)}{\left(r^2-2 r R \cos\theta +R^2\right)^{3/2}} = 0\\ \nonumber
	\dfrac{R \sin\theta \mu_m}{\left(r^2 - 2 r R \cos\theta + R^2\right)^{3/2}} + 2 \dot{r} \left(\dot{\theta} + \omega\right) + r \, \ddot{\theta} = 0
\end{gather}

Since the Moon is fixed in the rotating frame o reference, its position relative to the trajectory of the transfer is better visualized in this frame.
Furthermore, the specific force due to the gravitational influence of the Moon shows no explicit dependence on time in the rotating frame, which facilitates the evaluations.



\section{Numerical results}\label{re}

\begin{figure}[htbp]\centering
	\includegraphics[scale=0.35]{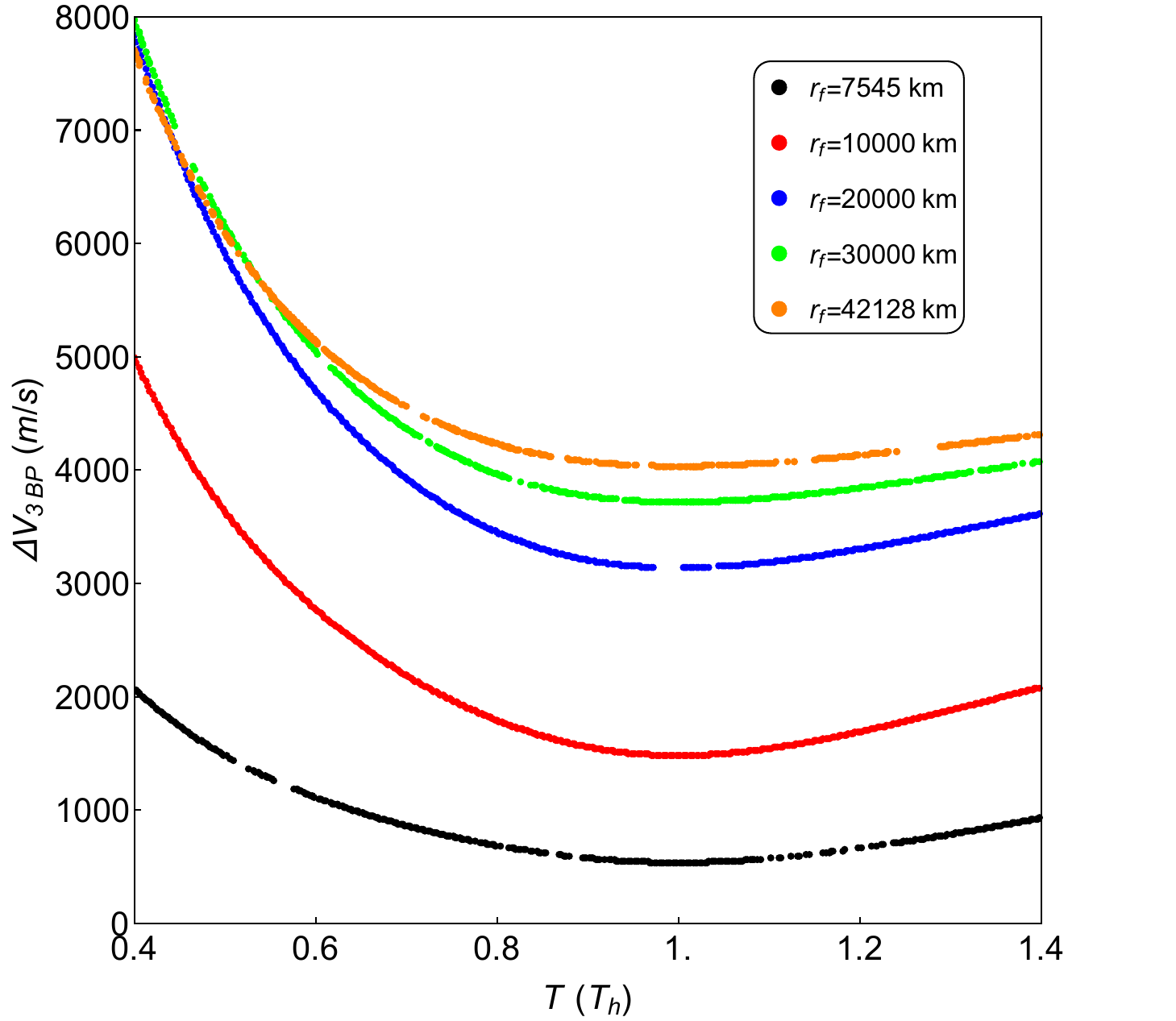}
	\caption{The costs of the one-tangent burn transfer as function of the time of flight (in units of the Hohmann transfer) for several values of the radius of the final orbit. The altitude of the initial orbit is $r_0 = 6,545$ km. $T_h$ is the time of flight of a Hohmann transfer between these orbits. Only counterclockwise ($\dot{\theta}(t)>0$) transfers are selected for this draw.}
	\label{fig:g1}
\end{figure}

\begin{figure}[htbp]\centering
	\includegraphics[scale=0.35]{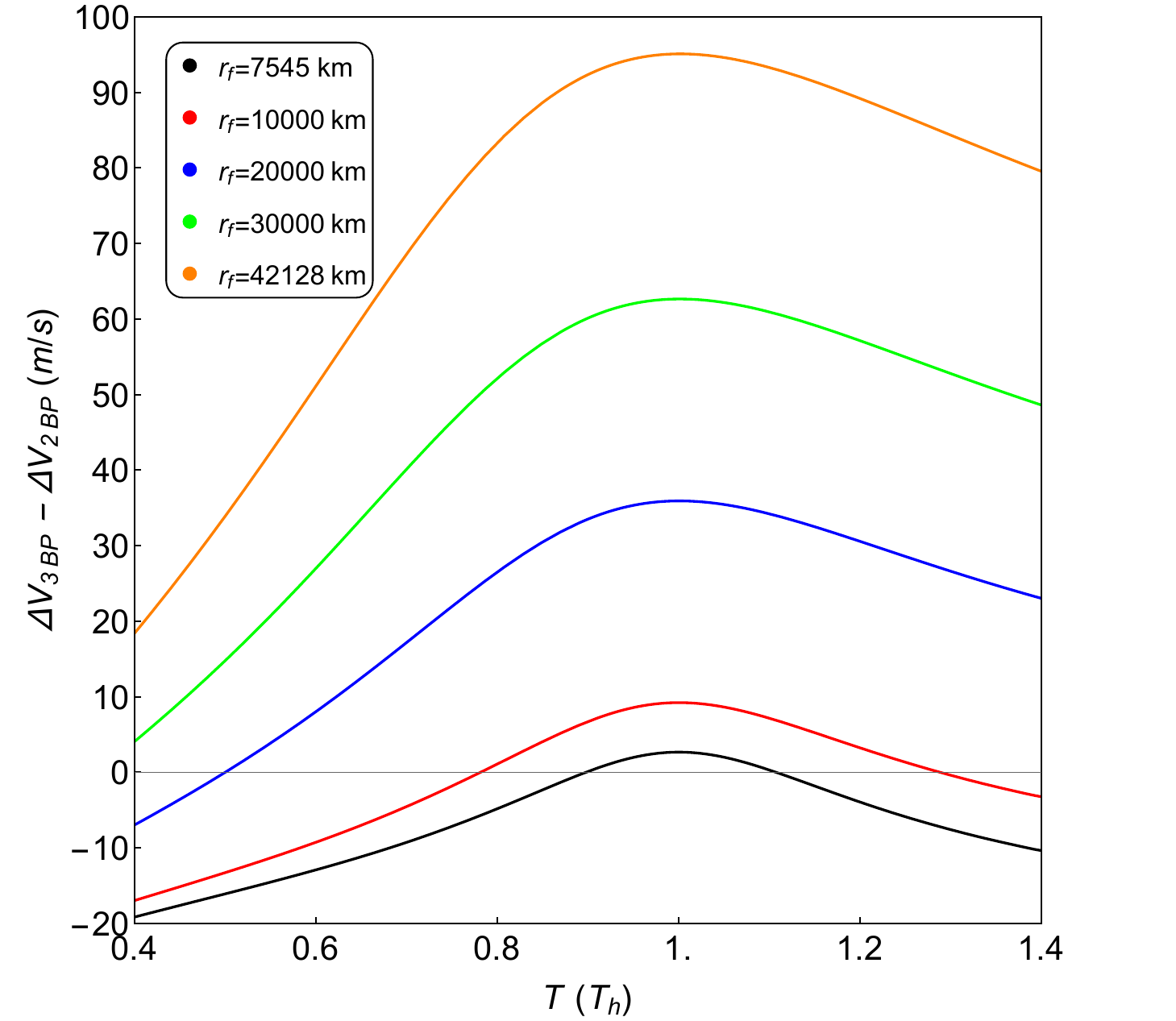}
	\caption{The differences of the costs between the CR3BP and the 2BP of the one-tangent burn transfer as function of the time of flight (in units of the Hohmann transfer) for several values of the radius of the final orbit. The initial orbit is $r_0 = 6,545$ km. $T_h$ is the time of flight of a Hohmann transfer between these orbits. Only counterclockwise ($\dot{\theta} (t) > 0$) transfers are selected for this draw.}
	\label{fig:g2}
\end{figure}

\begin{figure}[htbp]\centering
	\includegraphics[scale=0.4]{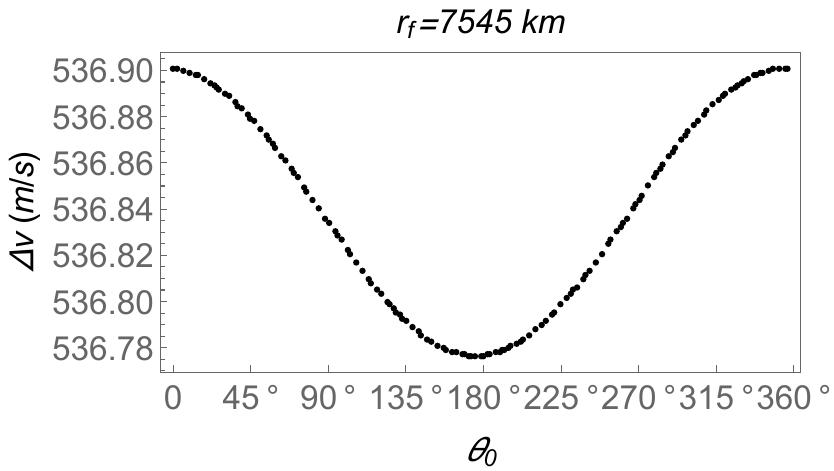}
	\includegraphics[scale=0.4]{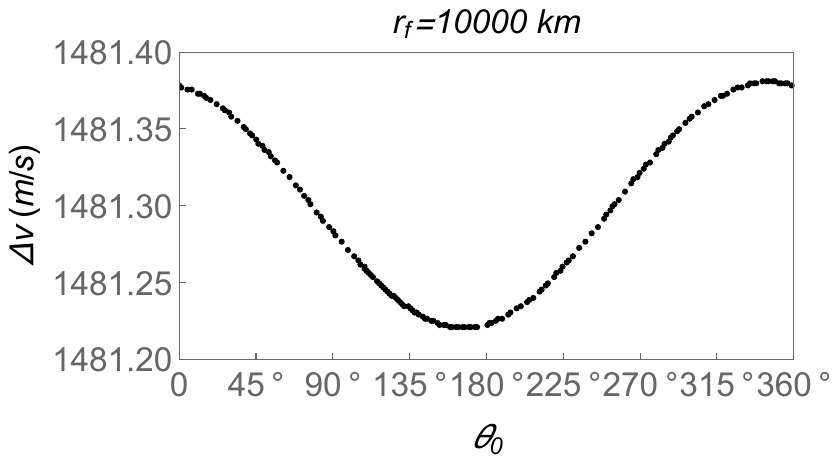}
	\includegraphics[scale=0.4]{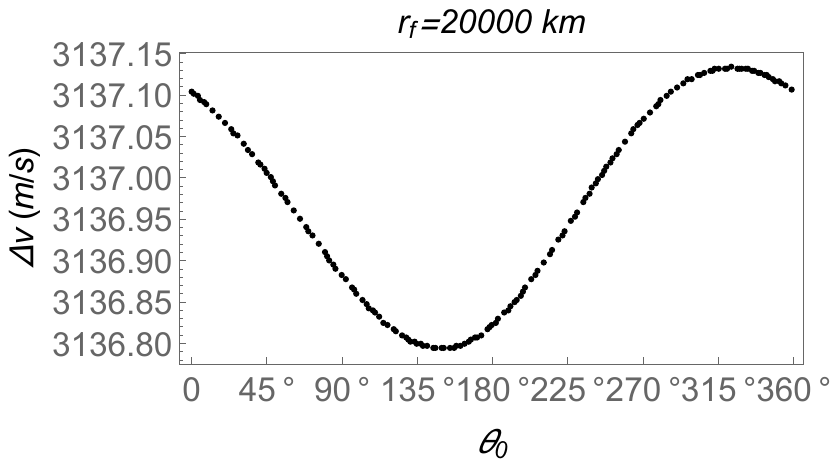}
	\includegraphics[scale=0.4]{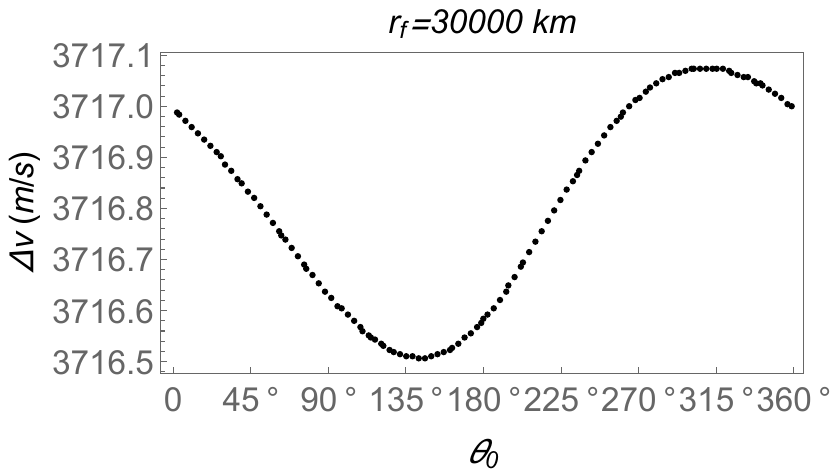}
	\includegraphics[scale=0.4]{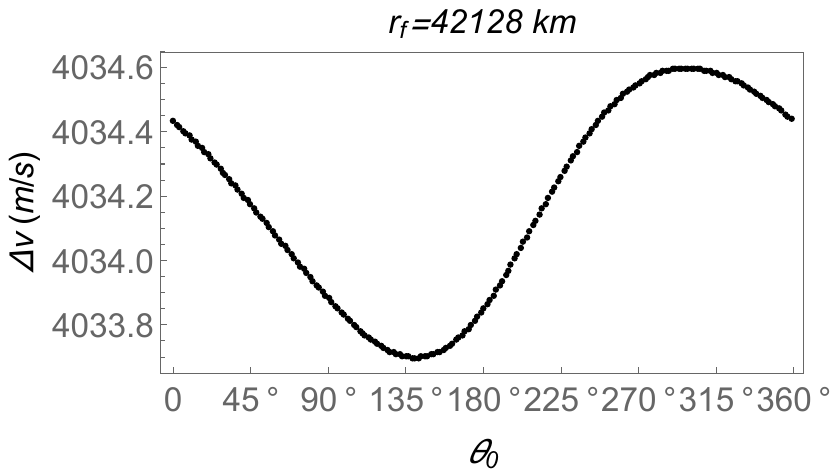}
	\caption{The costs for a one-tangent burn transfer as functions of the position of the first burn $\theta_0$ for several values of the radius of the final orbit $r_f$. The radius of the initial orbit is $R_e + 167$ km. The time of flight is $T = T_h$. Only counterclockwise ($\dot{\theta} (t) > 0$) transfers are selected for this draw.}
	\label{fig:dvw0}
\end{figure}

\begin{figure}[htbp]\centering
	\includegraphics[scale=0.4]{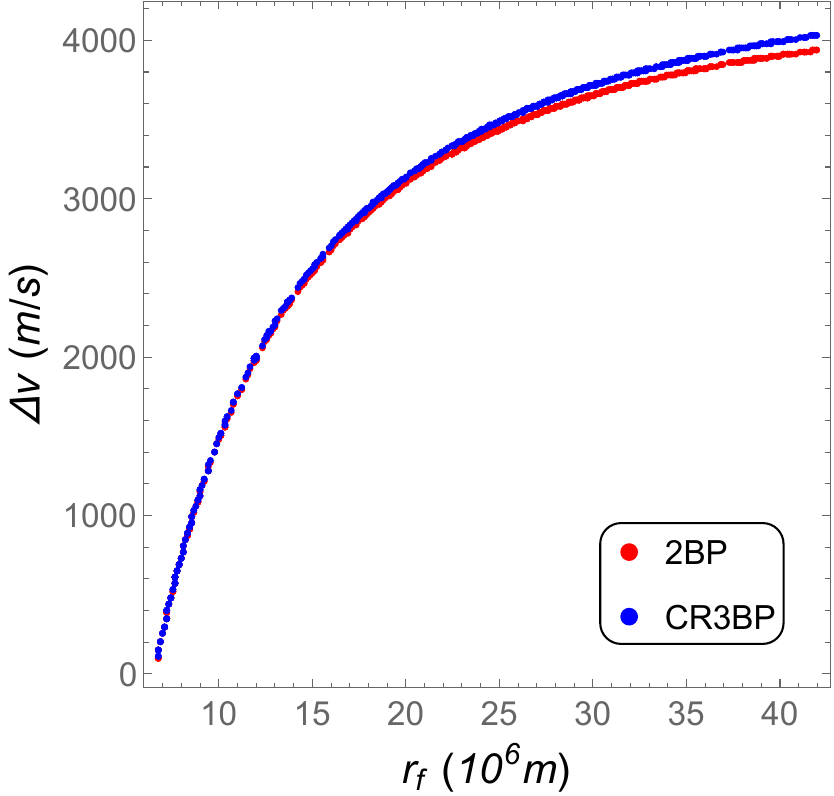}
	\includegraphics[scale=0.4]{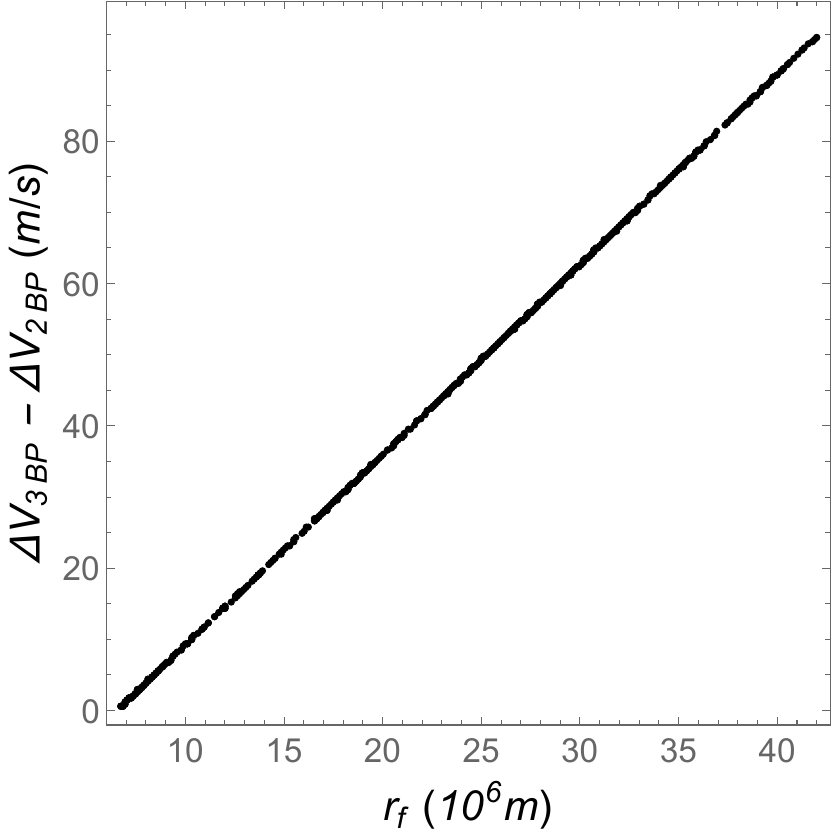}	
	\caption{The costs for the 2BP and the CR3BP (left) and their differences (right) for a one-tangent burn transfer as function of the radius of the final orbit. The radius of the initial orbit is $R_e + 167$ km. The first impulse is applied at a position such that $\theta (0) =-\pi/2$. Only counterclockwise ($\dot{\theta} (t) > 0$) transfers are selected in this draw.}
	\label{fig:dv23}
\end{figure}

\begin{figure}[htbp]\centering
	\includegraphics[scale=0.4]{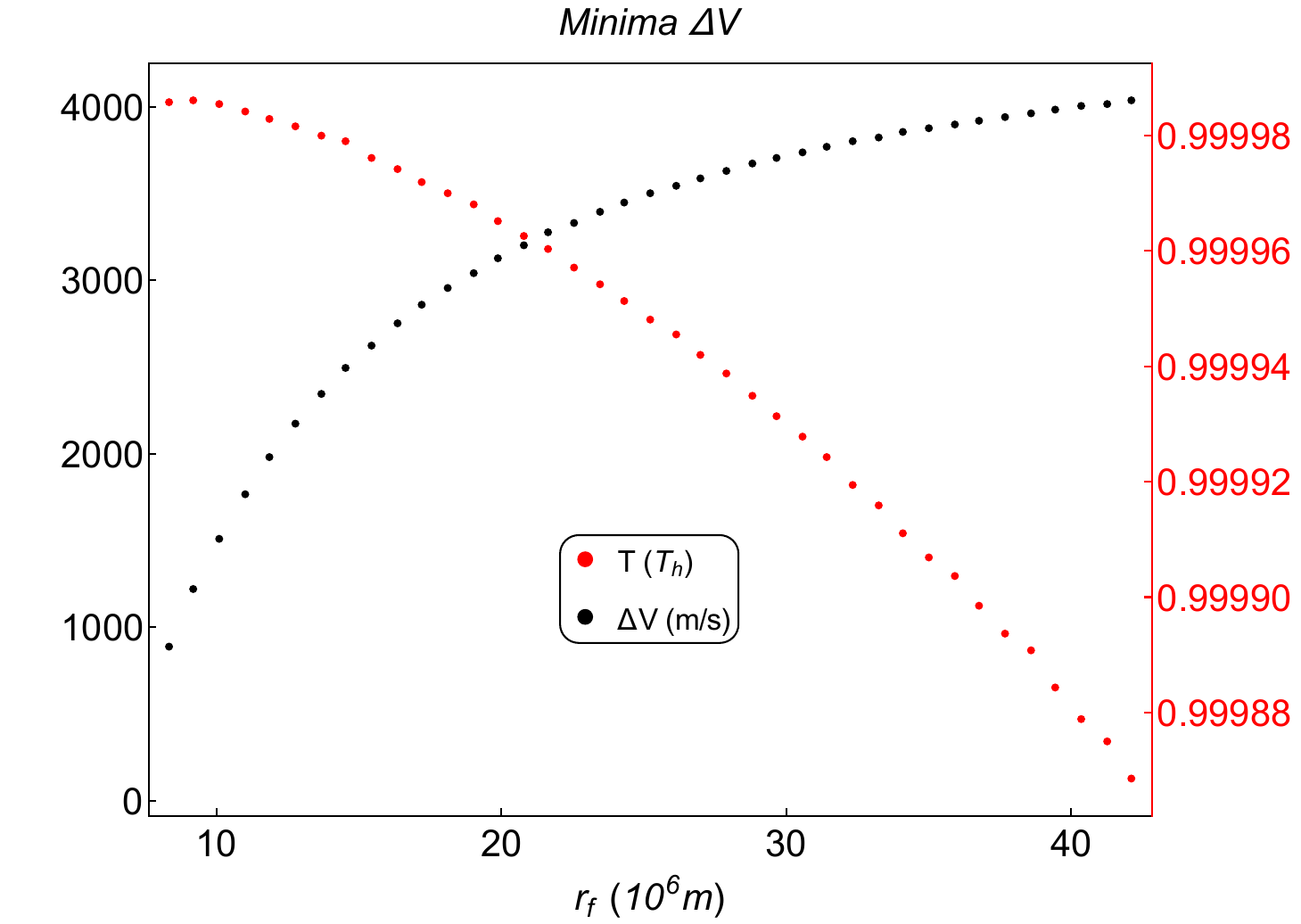}	
	\caption{The minimum costs $\Delta V$ and their associated time of flight as functions of the radius of the final orbit. The radius of the initial orbit is $R_e + 167$ km. The first impulse is applied at a position such that $\theta (0) =-\pi/2$. Only counterclockwise ($\dot{\theta} (t) > 0$) transfers are selected in this draw.}
	\label{fig:mindv}
\end{figure}

In this section, numerical evaluations are done for a transfer from a circular orbit of $167$ km of altitude (the radius of the initial orbit is $r_0 = R_e + 167$ km, where $R_e = 6,378$ km is the radius of the Earth) to another circular orbit whose altitude varies from $167$ km of altitude up to the altitude required for a geosynchronous orbit around the Earth ($r_f = 42,128.29441237582$ km).
The cost of the two-impulsive one-tangent burn transfer method is the magnitude of the difference between the velocity of the satellite in the initial circular orbit and the initial velocity of the transfer orbit plus the magnitude of the difference of the final velocity of the transfer orbit and the velocity of the final circular orbit.
These costs are shown in Fig. \ref{fig:g1} for several values of the radius of the final orbit as functions of the time of flight in units of $T_h$, which is the time of transfer of a Hohmann transfer between the respective circular orbits ($T=1$ means that the time of flight is equal to the time of a Hohmann transfer).
Using these units, the minimum of the costs of the transfer in the CR3BP can be clearly compared with the minimum of the costs under the 2BP, which is the time of a Hohmann transfer.
The minimum is very close to $T_h$, because the Moon exerts a small perturbative gravitational force over the spacecraft being transferred. The influence of the Moon over the costs of the transfers can be clearly seen in Fig. \ref{fig:g2}, through the difference of the costs between the CR3BP and the 2BP.
The closer this difference is to zero, the lower is the effect of the Moon over the costs.
Note that the gravity of the Moon is beneficial (it lower the costs) in some ranges of the time of flight, which vary for different values of the final orbit $r_f$. In general, the gravitational effect due to the influence of the Moon tends to increase the costs of the transfer when the time of flight is close to $T\approx T_h$ more than it tends to increase the costs for faster transfers. In fact, for transfers to orbits up to $r_f = 20,000$ km, the gravity of the Moon can lower the costs of the transfers - they are represented by the points where $\Delta V_{3BP}-\Delta V_{2BP}<0$.
The evaluations shown in Figs. \ref{fig:g1} and \ref{fig:g2} were done taking into consideration that the first impulse is applied at a point in the orbit such that $\theta_0 =-\pi/2$. On the other side, the costs do not have significant change for other values of $\theta_0$, as can be seen in Fig. \ref{fig:dvw0}, where the costs taking the gravity of the Moon into consideration are shown as functions of $\theta_0 = \theta (0)$. Note that, due to the symmetry of the problem, the position of the first impulse in the orbit has no influence over the costs for the 2BP.
The costs, as functions of the radius of the final orbit $r_f$, is shown in Fig. \ref{fig:dv23} (left), for a time of flight fixed in $T = T_h$. The differences of the costs between the CR3BP and the 2BP are shown in the right side of Fig. \ref{fig:dv23}. Note that these differences, as functions of the radius of the final orbit, are approximately linear.
Numerical evaluations were used to find the minimum costs and their associated values of the time of flight as functions of the radius of the final orbit, which are shown in Fig. \ref{fig:mindv}.

As stated in subsection \ref{tfc}, the Chebyshev orthogonal polynomials is chosen to compose the set of basis functions given by $\B{h}(t)$. The number of basis functions used to obtain the results was between 30 and 60. Note that the nonlinear least squares method is applied to solve the resulting differential Eqs.~(\ref{eq:movpol}) in polar coordinates by minimizing the residuals. Only numerical results in which the square root of the sum of the squares of the residuals is of the order of $10^{-10}$ m/s$^2$ or smaller are considered. The Legendre orthogonal polynomials was also tested, instead of the Chebyshev ones, and no significant differences in the results was obtained. For instance, the differences in the obtained $\Delta V$ obtained with Chebyshev and Legendre (with the same number of basis fuctions) polynomials were of the order of $10^{-8}$ m/s.

Note that the problems solved in this paper can also be addressed using other numerical techniques to solve the boundary value problem, like the shooting method with a initial guess based in the Lambert problem and increments in the velocity evaluated using a gradient method \cite{de1995transfer} or with Pareto optimal solutions using the direct transcription and shooting strategy \cite{topputo2013optimal}. The shooting method requires an integration of the initial position with a initial guess in the velocity. This velocity is varied after each step according to the encounter of the integrated trajectory with the final (desired) orbit. This integration (done at each step of the shooting method) can be evaluated, for instance, using TFC with the constraints given by the initial value problem (given by the initials position and velocity). The numerical evaluations done in this paper showed that the computational time required by the TFC method to solve the initial value problem (required at each step of the shooting method to update the initial velocity) is very similar to the computational time required by the TFC method to solve the related boundary value problem (whose constraints are given by the transfer), although the exacts computational times depend on the initial guesses. This is an expected result, since they have similar number of constraints. The shooting method usually requires many steps to satisfy a required accuracy - the number of steps depends on the required accuracy. Hence, in comparison, except in the case where the shooting method solves the problem in the first step (with a ``perfect'' initial guess), the approach via TFC with the constraints given by the boundary value problem is more efficient than the shooting method with iterative solutions obtained from the TFC with the constraints given by the initial value problem.

\section*{Conclusions}

This paper solves two astrodynamics transfer problems subject to nonlinear constraints using the TFC framework. In these transfer problems the constraints are typically nonlinear and coupled when expressed in the traditional rectangular coordinates. Via change of variables these nonlinear constraints are transformed into linear, condition required by the TFC framework to solve the associated boundary values problems. In addition to obtain linear constraints, this change of variable allows also to obtain decoupled equations. This results has been obtained by taking advantage of the symmetries of the problem. These features are also very useful to other approaches to solve the transfer problems.


The change from rectangular to polar coordinates allowed the use of the TFC framework to solve the one-tangent burn transfer method.
Note that the Hohmann and one-tangent burn transfer methods are based in the 2-body problem.
Using the TFC framework, the constraints generated by those methods can be embedded in a more complex system in which other perturbations are included. This may be easily done because the functional derived by TFC (the \textit{constrained expression}) does not depend on the equations of motions.
The gravitational influence of the Moon is included in the numerical evaluations done in this work. 




The change of coordinates to polar can benefit different methodologies to solve the orbit transfer problems. For instance, a Sun-centered frame of reference could be used to predict the transfer trajectories between different orbits by including other perturbations, as for instance, the solar radiation pressure or the gravitational perturbation of other bodies. In this paper, the transfer is done from a LEO to a geosynchronous orbit, but it could be done to a further orbit.
Another example of application is that it can be used to evaluate transfers for satellites close to Earth, also taking into consideration perturbations like the drag, other terms of the Earth gravitational potential, etc.

\section*{Data availability statement}
The manuscript has no associated data.

\begin{acknowledgements}
	This work was supported by FAPESP - S\~ao Paulo Research Foundation through grants 2018/07377-6 and 2016/24561-0 and National Council for the Improvement of Higher Education (CAPES). This work is also supported by the European Regional Development Fund (FEDER), through the Competitiveness and Internationalization Operational Programme (COMPETE 2020) of the Portugal 2020 framework [Project SmartGlow with Nr. 069733 (POCI-01-0247-FEDER-069733).
\end{acknowledgements}


\textbf{Data Availability Statement}: no Data associated in the manuscript.

\bibliographystyle{unsrt}
\bibliography{ref2}

\end{document}